\documentclass[]{aa} 

\usepackage{graphicx}
\usepackage{txfonts}
\usepackage{natbib}
\usepackage{longtable}
\usepackage{lscape}
\usepackage{color}

\providecommand{\sorthelp}[1]{}

\begin{document}
 
\title{
{LOC program for line radiative transfer}
}

\author{Mika  Juvela}

\institute{
Department of Physics, P.O.Box 64, FI-00014, University of Helsinki,
Finland, {\em mika.juvela@helsinki.fi}
}

\authorrunning{M. Juvela et al.}

\date{Received September 15, 1996; accepted March 16, 1997}

\abstract { 
{Radiative transfer modelling is part of many astrophysical simulations and is
used to make synthetic observations and to assist analysis of observations. In this
paper, we concentrate on the modelling of the radio lines emitted by the
interstellar medium. In connection with high-resolution models, this can be
significant computationally challenge.}
} 
{
Our aim is to provide a line radiative transfer (RT) program that makes good
use of multi-core central processing units (CPUs) and graphics processing
units (GPUs). Parallelisation is essential to speed up computations and to
enable the tackling of large modelling tasks with personal computers.
}
{
The program LOC is based on ray-tracing (i.e. not Monte Carlo) and uses
standard accelerated lambda iteration (ALI) methods for faster convergence.
The program works on 1D and 3D grids. The 1D version makes use of symmetries
to speed up the RT calculations. The 3D version works with octree grids and,
to enable calculations with large models, is optimised for low memory usage.
}
{
Tests show that LOC gives results that are in agreement with other RT codes
to within $\sim$2\%. This is typical of code-to-code differences, which often
are related to different interpretations of the model set-up. LOC run times
compare favourably especially with those of Monte Carlo codes. In 1D tests,
LOC runs were by up to a factor $\sim$20 faster on a GPU than on a single CPU
core. In spite of the complex path calculations, up to $\sim$10 speed-up was
observed also for 3D models using octree discretisation. GPUs enable
calculations of models with hundreds of millions of cells, as encountered in
the context of large-scale simulations of interstellar clouds.
}
{ 
LOC shows good performance and accuracy and and is able to handle many RT
modelling tasks on personal computers. Being written in Python, with only the
computing-intensive parts implemented as compiled OpenCL kernels, it can also
a serve as a platform for further experimentation with alternative RT
implementation details.
}


\maketitle

\section{Introduction} \label{sect:intro}

Our knowledge of astronomical sources is based mainly on observations of the
radiation that is produced, removed, or reprocessed by the sources. This
applies both to continuum radiation and the spectral lines.  The sensitivity
to the local physical conditions and the velocity information available via
the Doppler shifts makes spectral lines a particularly powerful tool. The
interpretation of the observations is complicated by the fact that we can view
the sources only from a single direction. In reality, the sources are
three-dimensional objects with complex variations of density, temperature,
chemical abundances, and velocities.

Radiative transfer (RT) models help to understand the relationships between
the physical conditions in a source and the properties of the observed
radiation. In forward modelling, the given initial conditions lead to a
prediction of the observable radiation that is unique, apart from the
numerical uncertainties of the calculations themselves. The main challenges
are related to the large size of the models (in terms of the number of volume
elements) which may call for simplified RT methods, especially when RT is
coupled to the fluid dynamics simulations. The model size can be a problem
also in the post-processing of simulations, especially if more detailed RT
calculations are called for. As a result, one may again have to resort to
supercomputer-level resources.

In the inverse problem, when one is searching for a physically plausible model for
a given set of observations, the models are usually smaller. However, modern
observations can cover a wide range of dynamical scales, thus setting corresponding
requirements on the models. The main problems are connected with the large
parameter space of possible models. It is difficult to find any model that matches
the observations, or the fact that the problem is inherently ill-posed means that
(within observational uncertainties) there may be many possible solutions for which
the allowed parameter ranges need to be determined. Therefore, the RT modelling of
a set of observations (the inverse problem) can be just as time-consuming and
computationally demanding as the post-processing of large-scale simulations (the
forward problem). Because of this complexity, observations are still often
analysed in terms of spherically symmetric models, although 2D and 3D modelling is
becoming more common. This is not necessarily bad, forced upon us by the
computational cost, but can serve as a useful regularisation of the complex
problem.

There is already a number of freely-available RT programs, some of which were
compared in \citet{Zadelhoff2002} and \citet{Iliev2009}, and new programs and new
versions of established codes continue to appear \citep{Olsen2018}. Within
interstellar medium studies, the codes  used in the modelling of radio spectral
lines range from programs using simple escape probability formalism, such as RADEX
\citep{van_der_Tak_2007}, via 1D-2D codes like RATRAN \citep{Hogerheijde2000} or
ART \citep{Pavlyuchenkov2004} to programs using full 3D spatial discretisation,
often in the form form of adaptive or hierarchical grids. As examples, MOLLIE
\citep{Keto2010} uses Cartesian and nested grids, LIME \citep{Brinch2010}
unstructured grids, ART$^3$ \citep{Li2020_ART2} octree grids, RADMC-3D
\citep{Dullemond2012} patch-based and octree grids, and Magritte
\citep{DeCeuster2020} even more generic information about the cell locations.
Programs are still often based on Monte Carlo simulations -- above only MOLLIE and
Magritte appear to use deterministic (non-random) ray tracing. In 3D models and
especially when some form of adaptive or hierarchical grids is used, it is
essential for good performance that the RT scheme takes the spatial discretisation
into account. Otherwise, information of the radiation field cannot be transmitted
efficiently to every volume element. Cosmological simulations deal with analogous
problems and various ray-splitting schemes are used to couple cells with the
radiation from discrete sources \citep{Razoumov2005, Rijkhorst2006,
Buntemeier2016}. This becomes more difficult in Monte Carlo methods, when the
sampling is random.

On modern computer systems, good parallelisation is essential. After all,
even on a single computer, the hardware may be capable of running tens (in
the case of central processing units, CPUs) or even thousands (in the case of
graphics processing units, GPUs) of parallel threads. Some of the programs
already combine local parallelisation (e.g. using OpenMP
\footnote{https://www.openmp.org/}) with parallelisation between computer
nodes (typically using MPI \footnote{https://www.open-mpi.org/}). In spite of
the promise of theoretically very high floating point performance of the
GPUs, these are not yet common in RT calculations and, within RT, are more
common in other than spectral line calculations \citep{Heymann2012,
Malik2017, Hartley2019, Juvela_2019_SOC}. However, parallel calculations over
hundreds or thousands of spectral channels should be particularly well suited
for GPUs \citet{DeCeuster2020}.

In this paper we present LOC, a new suite of radiative transfer programmes for
the modelling of spectral lines in 1D and 3D geometries, using deterministic ray
tracing and accelerated lambda iterations \citep{Rybicki1991}. The generated
rays follow closely the spatial discretisation, which is implemented using
octrees. LOC is parallelised with OpenCL
libraries\footnote{https://www.khronos.org/opencl/}, making it possible to run
the program on both CPUs and GPUs. LOC is intended to be used on desktop
computers, with the goal of enabling the processing of even large models with up
to hundreds of millions of volume elements. The program includes options for the
handling of hyperfine structure lines. There are also experimental routines
to deal with the general line overlap in the velocity space and the effects from
continuum emission and absorption. These features are not discussed in the
present paper that concentrates on the performance of LOC in the basic
scenarios, without the dust coupling and, in the case of hyperfine spectra,
assuming LTE distribution between the hyperfine components.

The contents of the paper are the following. The implementation of the LOC
programme is described in Sect.~\ref{sect:methods} and its performance is examined
in Sect.~\ref{sect:tests}, in terms of the computational efficiency and the
precision in selected benchmark problems. The technical details are discussed
further in Appendix~\ref{app:performance}. As a more practical example, we discuss
in Sect.~\ref{sect:example} molecular line emission computed for a large-scale
simulation of the interstellar medium (ISM). The results are compared with LTE
predictions and with synthetic dust continuum maps. We discuss the findings and
present our conclusions in Sect.~\ref{sect:discussion}.

\section{Implementation of the LOC programme}    \label{sect:methods}

LOC (acronym for line transfer with OpenCL) is a line radiative transfer program
that is based on the tracing of a predetermined set of rays through the model
volume\footnote{The program documentation can be found at
http://www.interstellarmedium.org/radiative\_transfer/loc/, with a link to the
source code at GitHub}. Unlike in Monte Carlo implementations, the sampling of the
radiation field contains no stochastic components.

The calculations use long characteristics that start at the model boundary and
extend uninterrupted through the model volume. The only exception are the rays
that in the 3D models are created in refined regions - these are discussed in
Sect.~\ref{sect:3D}. The rays express the radiation field in terms of photons
rather than intensity, an approach more typical for Monte Carlo methods.
However, LOC uses a fixed set of rays, without any randomness, and therefore is
not a Monte Carlo programme. Each ray contains a vector of photon numbers that
corresponds to a regular velocity discretisation of the spectral line profile
(typically some hundreds of channels). When a ray first enters the model, it carries
information of the background radiation field. This is expressed as the number
of photons that enter the model within each spectral channel
in one second.  If the background has an intensity of $I_{\nu}^{\rm bg}(\theta,
\phi)$ and there is one ray for a given surface area $A$ and solid angle
$\Omega$, the number of photons per channel is
\begin{equation}
n^{\nu} =  \frac{A \Delta v}{hc}  \int_{\Omega} I_{\nu}^{\rm bg}(\theta, \phi) \cos \theta d\Omega',
\end{equation}
where $\Delta v$ is the channel width in velocity, $h$ the Planck constant, and
$c$ the speed of light. The total number of photons coming from different
directions is represented with a small number of actual ray directions (in 3D
models typically some tens), but the total number of simulated photons is exact.
The number of photons per ray depends on the number of ray directions
(determining $\Omega$) and the number of rays per direction (determining $A$).
If the rays are equidistant in 3D space, $A$ still varies depending on the angle
between the ray direction and the local normal direction of the model surface.
If the rays are not equidistant, this causes additional changes in $A$ that are
taken into account in the above calculation. Non-equidistant rays are already
used for the 1D models (see Sect.~\ref{sect:1D}).

As a ray passes through a cell, we count the number of resulting induced upward
transitions $S_{lu}$ (from a lower level $l$ to an upper level $u$) in the
examined species. One must take into account both the photons that entered the
cell from outside as well the photons that were emitted within the cell,
\begin{equation}
S_{lu} = \frac{h \nu s}{4 \pi V}  B_{lu} \phi(v) 
\left( n_{\nu,0} \frac{1-e^{-\tau_{\nu}}}{\tau_{\nu}} 
+ n_{u} A_{ul} V \phi(\nu) \left[    1- \frac{1-e^{-\tau_{\nu}}}{\tau_{\nu}}
\right] \right),
\label{eq:Slu}
\end{equation}
\citep[cf.][]{Juvela1997}.
Here $s$ is the distance travelled across the cell, $\tau$ the corresponding
optical depth in a spectral channel, $V$ the cell volume, $n_{\nu, 0}$ the number
of incoming photons, $A_{ul}$ the Einstein coefficient for spontaneous emission, and
$\phi(\nu)$ the line profile function. The total number of upward transitions is
the sum over all spectral channels. The latter term within the parentheses
represents the photons that cause new excitations in the cell where they are
emitted. The number of photons within the ray changes similarly because of the
absorptions and emissions and, at the point where the ray leaves the cell, is equal
to
\begin{equation}
n_{\nu} = n_{\nu,0}e^{-\tau_{\nu}} + n_u A_{ul} V
\frac{1-e^{-\tau_{\nu}}}{\tau_{\nu}} \phi(\nu).
\label{eq:nnu}
\end{equation}

In practical calculations, $\phi(\nu)$ is replaced by the fraction of emitted
photons per channel. Furthermore, in both Eqs.~(\ref{eq:Slu}) and (\ref{eq:nnu})
one must take into account that each cell is traversed by a number of different
rays. Each ray is therefore associated only with a fraction of emission events that
is proportional to the length $s$. This can be calculated because the simulations
are deterministic and the total length $\Sigma s$ is known.

Equation~(\ref{eq:Slu}) results in estimates of the transition rates as an
effective average over the cell volume, as sampled by the rays, rather than
calculating intensities at discrete grid positions. This is again reminiscent of
the way the radiation-matter interactions are typically handled in Monte Carlo
RT programmes.

The radiation field computation is alternated with the solution of the equilibrium
equations that gives updated estimates of the level populations in each cell.  
There $S_{lu}$ is used directly in place of the normal rate of radiative
excitations, $B_{lu} I_{\nu}$, $B_{lu}$ being the Einstein $B$ coefficient. These
two steps are iterated until the level populations converge to required precision.
The final state is saved and spectral line maps are calculated for selected
transitions by a final line-of-sight (LOS) integration through the model volume.

LOC is parallelised using OpenCL libraries. The program consists of the main
program (written in Python) and a set of OpenCL kernels (written in the C
language) that are used for the computationally heavy tasks. In OpenCL
parlance, the main programme runs on the ``host'' and the OpenCL kernels are
run on the ``device'', which could be either the same main computer as for the
host (i.e. the kernel running on the CPU), or a GPU or another so-called
accelerator device. The simultaneous use of several devices is in principle
possible but is not yet implemented in LOC. Whether run on a CPU or a GPU, the
goal is to fully utilise the computing hardware for parallel RT calculations.
Parallelisation follows naturally from the parallel processing of the
individual rays and the parallel solving of the equilibrium equations for
different cells. In OpenCL, individual threads are called work items. A group
of work items forms a work group that executes the same instructions (on
different data) in lock-step. This sets some challenges for the load balancing
between the threads and synchronisation is possible only between the work
items of the same work group. GPUs allow parallel execution with a large
number of work groups, with even thousands of work items.

To speed up the convergence of the level populations for optically thick models,
LOC uses accelerated lambda iterations (ALI). This is implemented in its simplest
form, taking into account the self-coupling caused by photons being absorbed in the
cell where they were emitted (the so-called diagonal lambda operator).
In practice this is done by excluding the latter term in Eq.~(\ref{eq:Slu})
and by scaling the Einstein $A_{ul}$ coefficients in the equilibrium equations with
the photons escape probability.
The use of ALI does not have a noticeable effect on the time it takes to process a
given number of rays but it requires additional storage, one floating point number
per cell to store the escape probabilities. It is also possible to run LOC
without ALI. LOC is optimised for low memory usage, especially because the amount
of GPU memory can be limited. The main design decision resulting from this is that
the RT calculations are executed one transition at a time. This reduces the memory
requirements but has some negative impact on the run times, because the same ray
tracing is repeated for each transition separately.

The ray tracing, including the calculation of the radiative interactions, is
implemented as an OpenCL kernel function. Similarly, the solving of the
equilibrium equations and the computation of the final spectral line maps are
handled by separate kernels. For some kernel routines there are also
alternative versions, e.g. to handle corresponding calculations in the case of
hyperfine structure lines. LOC consists of actually two host-side programmes,
one for 1D and one for 3D geometries. These are described below, as far as is
needed to understand their performance and limitations.

\subsection{LOC in 1D} \label{sect:1D}

The 1D model is spherically symmetric and consists of co-centric shells with
values for the volume density, kinetic temperature, amount of
micro-turbulence, the velocity along the radial direction, and the fractional
abundance of the species. It is not possible to include any rotation in the 1D
models. The radiation field is integrated along a set of rays with different
impact parameters. These are by default equidistant, but can be set according
to a power-law. This may be necessary if the innermost shells are very small
compared to the model size and could not otherwise be sampled properly without
a very large total number of rays. The distance that a ray travels through
each shell is pre-calculated. This speeds up the ray-tracing kernel that is
responsible for following the rays and counting the radiative interactions in
the cells. The 1D version of LOC includes the options for the handling of
hyperfine lines, with the LTE assumption or with the general line overlap, as
well as the effects of additional continuum emission and absorption
\citep[cf.][]{Keto2010}.

The 1D version of LOC processes each ray using a separate thread or, in OpenCL
parlance, a separate work item. Even in detailed 1D models, only up to some
hundreds of rays is needed. This suggests that 1D calculations will not make
full use of GPU hardware that might be able to provide many more hardware
threads. Nevertheless, in practical tests GPUs often outperformed CPUs by a
factor of few (Appendix~\ref{app:performance}).

\subsection{LOC in 3D} \label{sect:3D}

The spatial discretisation of the 3D LOC models is based on octree grids,
which in principle include regular Cartesian grids as a special case. In
practice, the Cartesian grid case is handled by a separate kernel, where the
ray-tracing algorithm is simpler and each ray is processed by a single work
item. On octree grids, the ray tracing is more complex and the computations on
a given ray are shared between the work items of a work group. The directions
of the rays are calculated based on Healpix pixelisation \citep{Gorski2005},
to ensure uniform distribution over the unit sphere. 

The 3D ray-tracing is based on two principles. First, for any given direction
of the rays, the total length of the ray paths is the same in all cells, apart from
the dependence on the cell size. There is thus no random variation in the radiation
field sampling. Second, we assume that we have no knowledge of the radiation field
variations at scales below the local spatial discretisation. This simplifies the
creation of new rays by avoiding interpolation. Especially when LOC is run on a
GPU, it would be costly (in terms of additional computations and synchronisation
overheads) to try to combine the information carried by different rays.

The calculations loop over the different ray directions. When a direction is
selected, the vector component with the largest absolute value determines the
``main direction'' of ray propagation. This tells which side of the model is most
perpendicular to the main direction and thus most illuminated by the radiation
with the current ray direction. This side is referred to as the upstream side or
border, both for the model and for the individual cells. One ray is started
corresponding to each of the surface elements on the upstream border of the model.
The step between the rays corresponds initially to the cell size at the root level
of the grid hierarchy. In the following, the level of the hierarchy level $L$=0
refers to the root grid, $L$=1 to the first subdivision to eight sub-cells, an so
forth. The initial number of concurrent rays is thus equal to the number of $L=0$
surface elements on the upstream border. The exact position of the rays within its
surface element is not important.

\begin{figure}
\includegraphics[width=8.4cm]{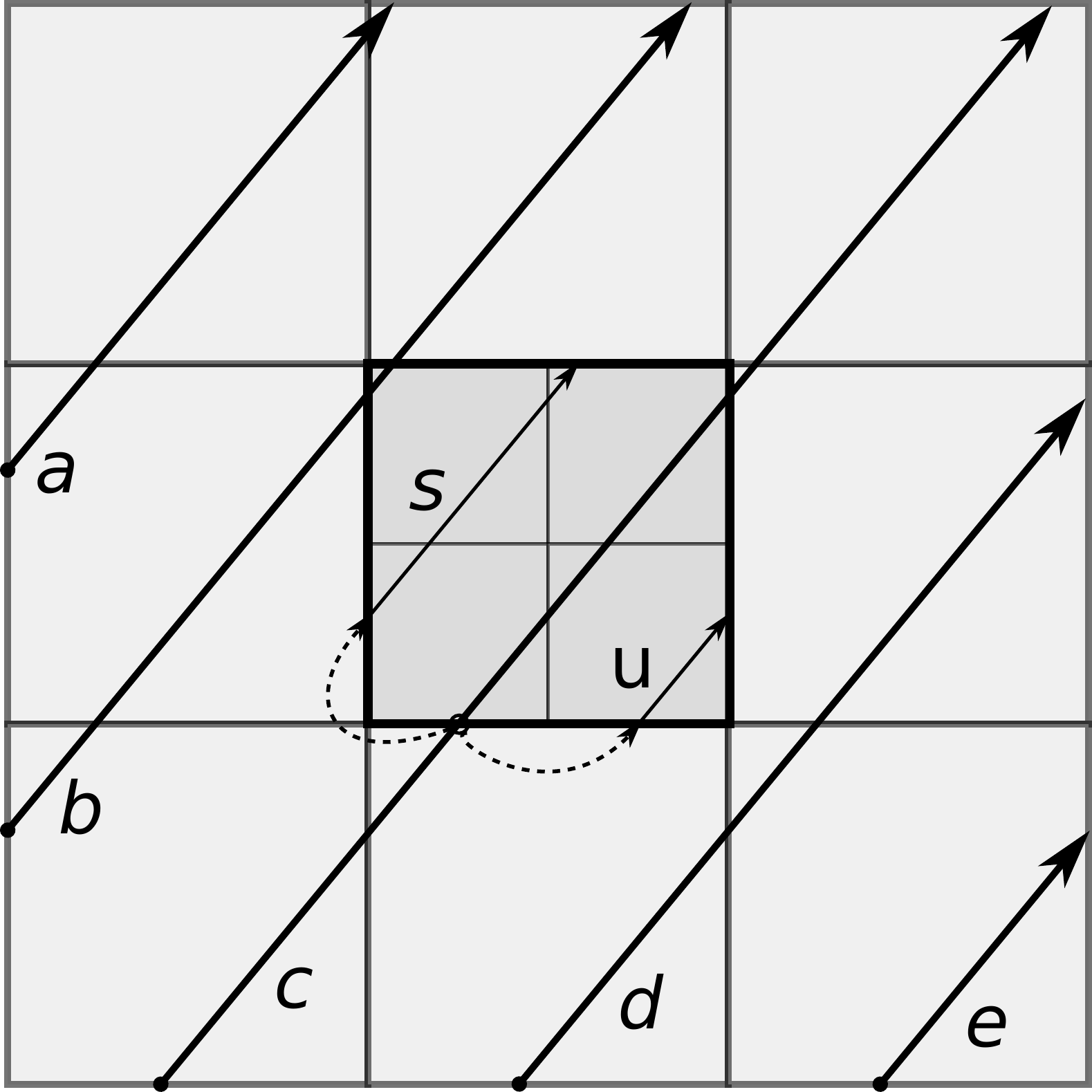}
\caption{
Illustration of the creation of rays for a refined region. Rays $a$-$e$ give a
uniform sampling at the coarser level. The rays $s$ and $u$ are added for the
refined region, using the information that ray $c$ has of the radiation field at the
upstream boundary of the refined cell.
}
\label{fig:drawing1}
\end{figure}

As rays are propagated through the model volume, these can enter regions with
higher or lower level of refinement. The grid of rays follows the grid of spatial
discretisation. When a ray comes from a cell at level $L$ to a cell $C$ that has
been split further to eight smaller level $L+1$ cells, and when that ray enters on
the upstream border of the cell $C$, new rays are added (cf.
Fig.~\ref{fig:drawing1}). The original ray enters
one of the level $L+1$ sub-cells of $C$, and three new rays are added
corresponding to the other level $L+1$ sub-cells of $C$ that share the same
upstream border. All four upstream rays (the original ray included) are assigned
one quarter of the photons of the original ray. 

To keep the sampling of the radiation field uniform at all refinement levels,
it may be necessary to add rays also on the other sides of the cell $C$ that
are perpendicular to the upstream border, two of which may also be illuminated
by radiation from the current direction. A ray needs to be created if a ray
corresponding to the discretisation level $L+1$ would enter $C$ through a
neighbouring cell that itself is not refined to the level $L+1$. If the
neighbour is at some level $L'<L$, all the rays corresponding to the
refinement at and below $L'$ will at some point exist in the neighbouring cell
and will be automatically followed into the cell $C$. These need not be
explicitly created. However, the side-rays that correspond to discretisation
levels above $L'$ and up to $L+1$ have to be added, using the information from
the original ray at the upstream border. If the step is $\Delta L=1$, the
number of photons in the newly created ray is one quarter of the photons
reaching the upstream cell boundary, as was the case for the new rays on the
upstream border. This is not double counting, since those photons should have
entered cell $C$ from a neighbouring cell, along a ray that did not exist
because the lower discretisation of the neighbouring cell. Unlike in the case
of new rays on upstream border, the need to add rays on the other sides has to
be checked one every step, not only when the refinement changes along the
original ray path.

The addition of the side-rays constitutes the longest extrapolation of the
radiation field information, the ray entering the upstream side of $C$ also
providing part of the information for the radiation field at some other sides
of $C$. Since side-rays are added only when the neighbouring cell is not
refined to the level $L+1$, the extrapolation is over a distance that
corresponds to the spatial discretisation, i.e. is less than the size of the
level $L$ cells. For example, in most ISM models, there is structure at all
scales and, if more accuracy is needed, it is usually better to improve the
discretisation than to rely on higher-order interpolation of non-smooth
functions.

Figure~\ref{fig:drawing1} illustrates the creation of new rays when the
discretisation level increases by one. The main direction is upwards and the thick
arrows correspond to the rays on the coarser discretisation level. The central
cell has been refined, the shading indicating the extent of the refined region.
When ray $c$ enters the refined region, one new ray $u$ is created on the upstream
boundary based on the ray $c$. In 3D, this would correspond to three new rays. This
can be contrasted with Fig. 2 of \citet{Juvela2005_bingrid}, where the creation of
a ray requires interpolation between three rays. Because in Fig.~\ref{fig:drawing1}
the neighbouring cell on the left is less refined than the cells in the refined
region, the ray $s$ has to be created at the boundary of the refined region. That
is based on information from the ray $c$. In Fig.~\ref{fig:drawing1}, also the ray
$b$ crosses the shaded refined cell. However, that ray already exists, because the
neighbouring cell on the left is refined to the same level as the central cell
(before it was refined once). If the neighbouring cell had been even less refined,
also the ray $b$ would have been created. That would have taken place at the point
where the ray $b$ enters the refined region and again only using information
carried by the ray $c$. In actual calculations, there are no guarantees on the
order in which the rays $a$-$e$ are processed. However, rays $s$ and $u$ will be
processed by the same work group as the ray $c$. Those rays will be followed until
they again reach a coarser grid and are thus terminated, and the computation of the
ray $c$ is resumed thereafter, at the point where it first entered the refined
region. The order is important in reducing the needed storage space (see below).

If the refinement level changes at a given cell boundary by more than $\Delta
L=1$, the above procedure is repeated recursively. Each of the new rays is
tagged with the hierarchy level where it was first created, in the above
example the level $L+1$. When a ray reaches a region that is refined only to
level $L$ or less, the ray is terminated. In theory, a better alternative
would be to join the photons of the terminated sub-rays back to the ray(s)
that continue into the less-refined region. However, this would again require
synchronisation and interpolation between rays that otherwise would not yet or
no longer exist (as they may be computed by other work groups, possibly during
entirely different kernel calls). However, the termination of rays and thus
the associated loss of information again only involves scales below the
resolution of the local discretisation. When the refinement level decreases,
the photon count of the rays that are not terminated is correspondingly scaled
up by $4^{-\Delta L}$, $\Delta L$ being the change in the refinement level
(here a negative number).

When a ray exits the model volume through some side other than the downstream
border, the corresponding work group (or, in the case of a Cartesian grid, a
work item) creates a new ray on the opposite side of the model, at the same
coordinate value along the main direction. The work group (work item) finishes
computations only when the downstream border of the entire model is reached.
This also helps with the load balancing, although different rays may of course
encounter different amounts of refined regions. With this ray-tracing scheme, 
each cell is traversed by at least by one ray per direction, and the physical
path length is the same in all cells, except for the 0.5$^L$ dependence on the
discretisation level.

One refinement requires the storage of three rays from the upstream border,
the original level $L$ ray and two new rays that only exist at levels $L+1$
and higher. The third of the new rays will be continued immediately.  At most
four rays may be created at the other sides of the cell $C$. When one ray is
terminated, the next ray is taken from the buffer, unless than is already
empty. To minimise the memory requirements, one always simulates first the
rays created at higher refinement levels. The splitting of a ray requires the
storage the locations of the rays and the single vector containing the
original number of photons per velocity channel. Since the splitting is
repeated for each increase in refinement level, one root-grid ray may lead to
a $\sim 7 \times 3^{N_{\rm L}-1}$ rays being stored a buffer. This can become
significant for large grids (e.g. a 512$^3$ root grid corresponding to 512$^2$
simultaneous root-level rays). However, if necessary, the number of concurrent
rays can be limited by simulating the root-grid rays in smaller batches.

The use of a regular grid of rays eliminates random errors in the sampling of
the radiation field. The remaining numerical noise in the path lengths per
cell is in LOC below 10$^{-4}$ (tested up to six octree levels) and is thus
mostly insignificant. While the path lengths are constant, the locations of
the rays are not identical in all cells. This is part of the unavoidable
sampling errors, but is again an error that is related to radiation field
variations at scales below the spatial discretisation. The number of angular
directions used in the calculations of this paper is 48.

The ray-tracing scheme used on octree grids, with the creation and termination
of rays, is more complex than the simple tracing of individual rays through an
octree hierarchy. Close to 90\% of the code in the kernel for the radiation
field simulation involves just the handling of the rays. As an alternative, it
would be possible to use brute force and simulate a regular grid of rays with
$4^{N_{\rm L}}$ rays for every level $L=0$ surface element. The simplified ray
tracing makes this competitive for grids with two levels, but it becomes
impractical for deeper hierarchies because of the $4^{N_{\rm L}}$ scaling.

\section{Test cases}    \label{sect:tests}

\subsection{Tests with one-dimensional models}  \label{sect:1d}

\begin{figure}
\includegraphics[width=8.4cm]{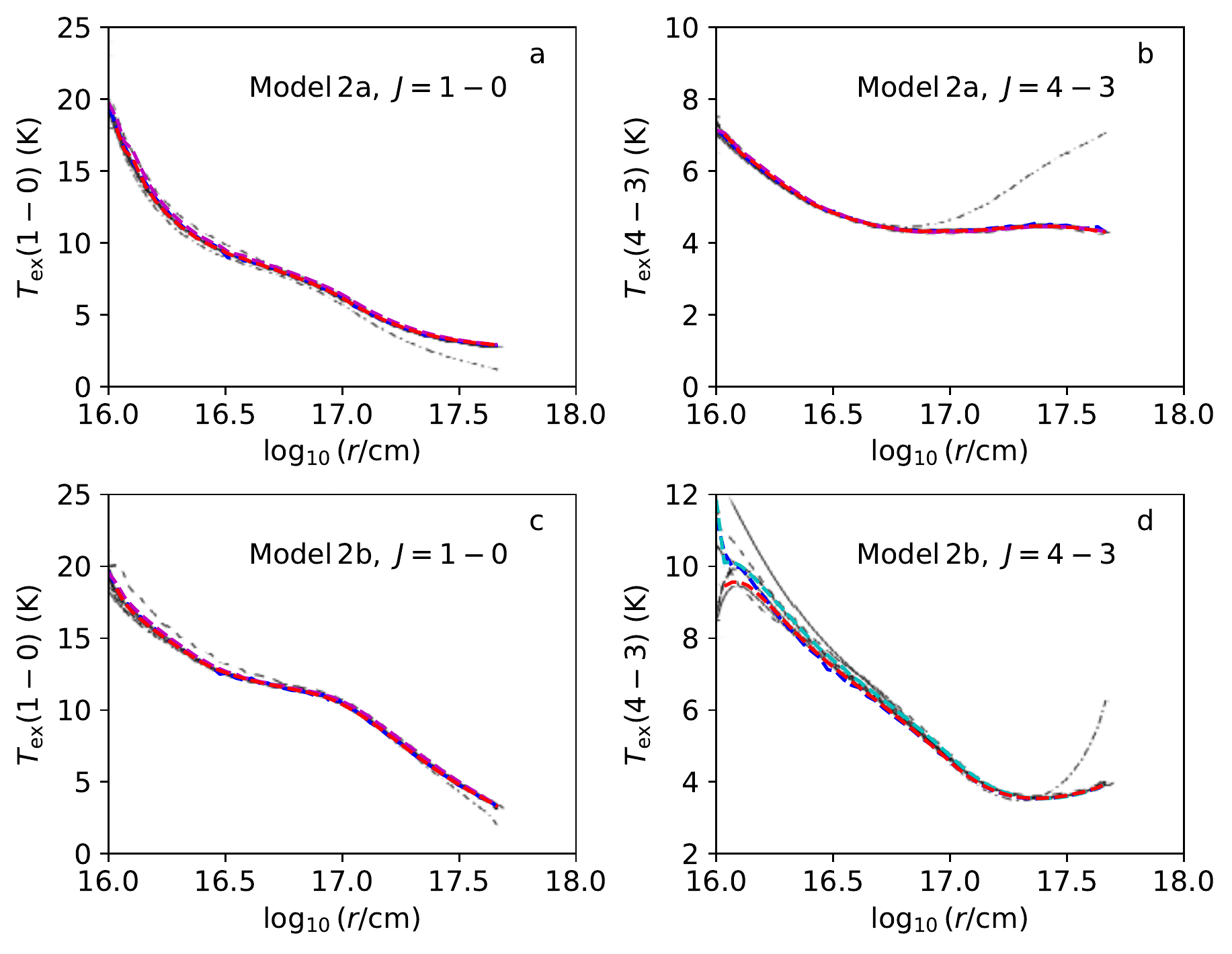}
\caption{
Comparison of 1D LOC results for cloud models presented in \citet{Zadelhoff2002}.
The plots show the radial excitation temperature profiles of the HCO$^+$ $J=1-0$
and $J=4-3$ transitions for Model 2a and the optically thicker Model 2b. The LOC
results are plotted with red hashed lines, the results of the Monte Carlo code
Cppsimu in blue, and the other results from \citep{Zadelhoff2002} in black (with
different line types). In frame $d$, the cyan line shows LOC results for an
alternative interpretation of the problem set-up.
}
\label{fig:model2}
\end{figure}

We test the 1D LOC first using the Models 2a and 2b from
\citet{Zadelhoff2002}. These are 1D models of a cloud core with infall motion,
with 50 logarithmically divided shells, and with predictions computed for
HCO$^+$ lines. The two cases differ only regarding the HCO$^+$ abundance that
is ten times higher in Model 2b, in that case increasing the optical depth to
$\tau\sim$4800. \citet{Zadelhoff2002} compared the results of eight radiative
transfer codes, Fig.~\ref{fig:model2} showing a reproduction of these. We
overplot in the figure new computations with the Monte Carlo programme
Cppsimu \citep{Juvela1997} and the 1D version of LOC.

The LOC results follow closely the average of the results reported in
\citet{Zadelhoff2002} as well as the recomputed Cppsimu $T_{\rm ex}$ curve.
Noticeable differences appear towards the centre of the optically thicker Model 2b,
where LOC gives some of the lowest $T_{\rm ex}$ values, although still within the
spread of the 2002 results. The test problem was specified using a grid of 50
radial points. The results plotted in red were computed assuming that the grid
points correspond to the outer radii of the shells and the listed physical values
refer to the shells inside those radii. Especially in the inner part model, it also
becomes important, how the codes deal with the velocity field. In LOC the Doppler
shifts are evaluated only at the beginning of each step. Alternatively, the
magnitude and direction of the velocity vector could be evaluated at the centre of
the step, or even for several sub-steps separately. The LOC calculations used
$\sim$100 velocity channels, the channel width being at most one fifth of the full
width at half maximum (FWHM) of the local profile functions in the cells.

To illustrate the sensitivity of the results to the actual model set-up, we
show in Fig.~\ref{fig:model2} results for an alternative LOC calculation. This
uses the same discretisation as above (only splitting the innermost cell to
two) but interpolate the input data to the radial distances that correspond to
the average of the inner and outer radius of each cell. This results in
changes that are of similar magnitude as the differences between the codes.
The $T_{\rm ex}$ values of the second LOC calculation even rise above the
Cppsimu values. As noted in \citet{Zadelhoff2002}, better discretisation tends
to decrease the differences between the codes. This reduces the effect of
assumptions made at the level of the model discretisation and shows that the
actual differences in the RT program implementations are only partly
responsible for the scatter in the results.

\subsection{Comparison of 1D and 3D models} \label{sect:1D3D}

To test the 3D version of LOC, we first compare it against 1D calculations made
with Cppsimu and LOC. We adopt a spherically symmetric model that has a radius of
$r=0.1$\,pc and the corresponding 3D models are discretised over a volume of
$0.2\times0.2\times0.2$\,pc. The density distribution has a Gaussian shape with a
centre density of $n({\rm H_2})=10^6$\,cm$^{-3}$ and a FWHM of 0.05\,pc. The model
has a radially linearly increasing infall velocity (0-1\,km\,s$^{-1}$), kinetic
temperature (10-20\,K), turbulent linewidth (0.1-0.3\,km\,s$^{-1}$), and fractional
abundance. The radial discretisation of the 1D model consist of 101 shells, with
the radii placed logarithmically between 10$^{-4}$ and 0.1\,pc. The 3D model
extends further towards the corners of the cubic volume but the hydrogen number
density at the distance of 0.1\,pc is already down to $n({\rm H_2})\sim
1.5$\,cm$^{-3}$. 

\begin{figure}
\includegraphics[width=8.4cm]{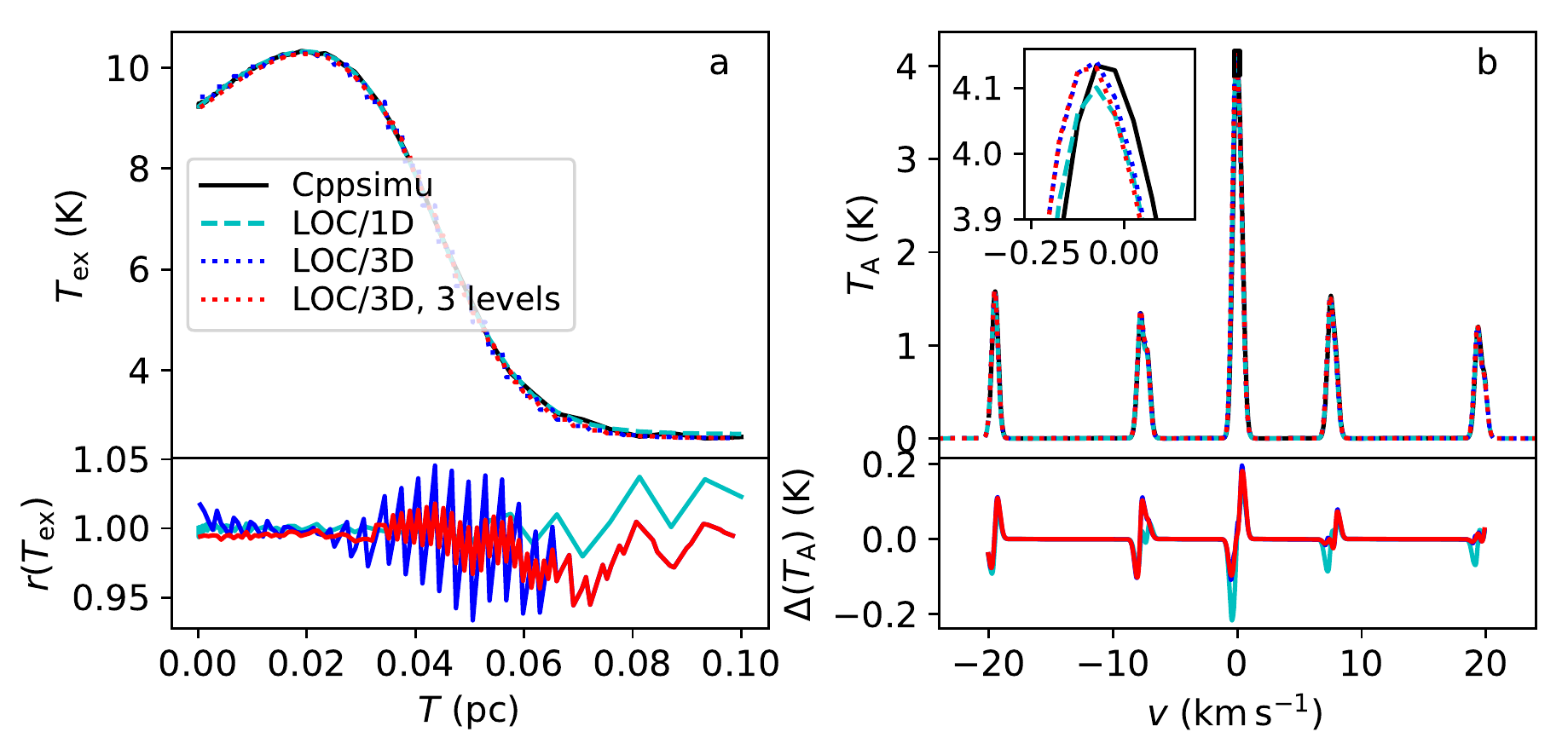}
\caption{
Comparison of NH$_3$(1,1) excitation temperatures and spectra for a spherically
symmetric cloud. Frame $a$ shows the radial $T_{\rm ex}$ profiles computed with
1D versions of the Cppsimu and LOC programmes and with the 3D version of LOC.
The LOC runs use either a Cartesian grid with 64$^3$ cells or the same with two
additional levels of refinement. Frame $b$ compares the spectra observed towards
the centre of the model. A zoom into the peak of the main component is shown in
the inset. The lower plots show the ratio of $T_{\rm ex}$ values (frame
$a$) and the difference between the LOC and Cppsimu $T_{\rm A}$ values (frame $b$),
with the colours listed in frame $a$.
}
\label{fig:NH3}
\end{figure}

Figure~\ref{fig:NH3} shows the results calculated for ammonia, including the
hyperfine structure for the shown NH$_3$(1,1) lines and assuming an LTE
distribution between the hyperfine components \citep{Keto2010}. The fractional
abundance is set to increase from 10$^{-8}$ in the centre to
2$\times$10$^{-8}$ at the distance of 0.1\,pc. Figure~\ref{fig:NH3} shows the
radial excitation temperature profiles and the spectra towards the centre of
the model. In addition to the 1D calculations performed with Cppsimu and LOC,
there are two sets of results from the 3D LOC. The first model uses a regular
Cartesian grid of 64$^3$ cells (with a size of 1.5625\,mpc per cell) and the
second adds two levels of octree refinement, giving an effective resolution of
0.39\,mpc. At each level, 15\% of the densest cells of the previous hierarchy
level are refined.

The excitation temperatures of the models match to within 5\%. The largest
differences result from the low spatial resolution of the pure Cartesian grid,
combined with the regions of the steepest $T_{\rm ex}$ gradients. The relative
differences in $T_{\rm ex}$ and absolute difference in $T_{\rm A}$ are shown
relative to the Cppsimu results, while the agreement between 1D and 3D LOC
versions is slightly better. For the spectra, the residuals (up to $\Delta
T_{\rm A}\sim 0.2$\,K) are mainly caused by small differences in the velocity
axis (a fraction of one channel, due to implementation details),
while the peak $T_{\rm A}$ values match to within 1\%.

\subsection{Large-scale ISM simulation} \label{sect:example}

As an example of a more realistic RT modelling application, we compute
molecular line emission from a large-scale MHD simulation of turbulent ISM. We
use this to examine the convergence of the calculations in the case of an
partially optically thick model, and to demonstrate the importance of non-LTE
excitation.

\subsubsection{Model setup}

The model is based on the magnetohydrodynamic (MHD) simulations of
supernova-driven turbulence that were discussed in \citet{Padoan2016} and were
previously used in the testing of the continuum RT program SOC
\citep{Juvela_2019_SOC}. The model covers a volume of (250\,pc)$^3$ with a mean
density of $n({\rm H})=5\,{\rm cm}^{-3}$.

We use LOC to calculate predictions for the $^{12}$CO(1-0) and $^{13}$CO(1-0)
lines, using the molecular parameters from the Leiden Lamda
database\footnote{https://home.strw.leidenuniv.nl/~moldata/} \citep{Schoier2005}.
Since the used MHD run did not provide abundance information, we assume for
$^{12}$CO(1-0) abundances an ad hoc density dependence
\begin{equation}
\chi = 10^{-4} \, n^{2.45} (3\times 10^8+n^{2.45})^{-1}.
\label{eq:abundance}
\end{equation}
This corresponds to the simulations of \citet{Glover_2010}. The fractional
abundance of $^{13}$CO molecule is assumed to be lower by a factor of 50.

Because of the large volume, the model contains a wide range of velocities but
we include in the calculations only a bandwidth of 50\,km\,s$^{-1}$.  The
larger velocities are exclusively associated with hot, low-density gas. With
the assumed density dependence, the CO abundance of these regions is
negligible. 

We run LOC on a system with where the host computer has only 16\,GB of main
memory. This requires some optimisation to reduce the number of cells in the
model. The original MHD data correspond to a root grid of 512$^3$ cells and six
levels of refinement. The total number of cells is 450 million. To reduce the
memory footprint, we limit the RT model to the first three hierarchy levels and
a maximum spatial resolution of 0.12\,pc. However, because most cells are on
the lower discretisation levels, this reduces the number of cells only by 22\%,
down to 350 million.

Because regions with low gas densities do not contribute significantly to the
molecular line emission, we can do further optimisation by eliminating some of
the refined cells. Whenever a cell has been divided to sub-cells that all have
a density below $n({\rm H}_2)$=20\,cm$^{-3}$, the refinement is omitted,
replacing the eight sub-cells with a single cell. This is justified by the very
low CO fractional abundance at this density (Eq.~\ref{eq:abundance}). To
increase the effect of this optimisation, the octree hierarchy is also first
modified to have a root grid of 256$^3$ cells and four instead of three
refinement levels. By joining the low-density cells, the number of individual
cells is reduced to 105 million, allowing the model to be run within the given
stringent memory limits. The reduction in the number of cells of course also
reduces the run times that are approximately proportional to the number of
individual cells (see Appendix~\ref{app:performance}). 

The kinetic temperature was left to a constant value of 15\,K. The velocity
dispersion in each cell was estimated from the velocities of the child cells
(making use of the full model with seven hierarchy levels). The velocities were
calculated similarly as density-weighted averages over the child cells. The
number of velocity channels was set to 256, giving a resolution of
0.19\,km\,s$^{-1}$. This is comparable to the smallest velocity dispersion in
individual cells (including the thermal broadening). The number of energy level
included in the calculations is 10 (the uppermost level $J=9$ being about
300\,K above the $J=0$ level).

\subsubsection{Convergence} \label{sect:convergence}

The rate of convergence, and thus the number of required iterations, depends on
the optical depths and will vary between regions. We are not affected by random
fluctuations that in Monte Carlo methods would make it more challenging to
track the convergence. Although $^{13}$CO can become optically thick in some of
the densest clumps, convergence should be clearly slower for the main isotopic
species. However, because of the strong inhomogeneity of the model, even the
$^{12}$CO(1-0) line is optically thick only for a couple of percent of the LOS,
in spite of the maximum optical depth being close to $\tau=100$.

\begin{figure}
\includegraphics[width=8.8cm]{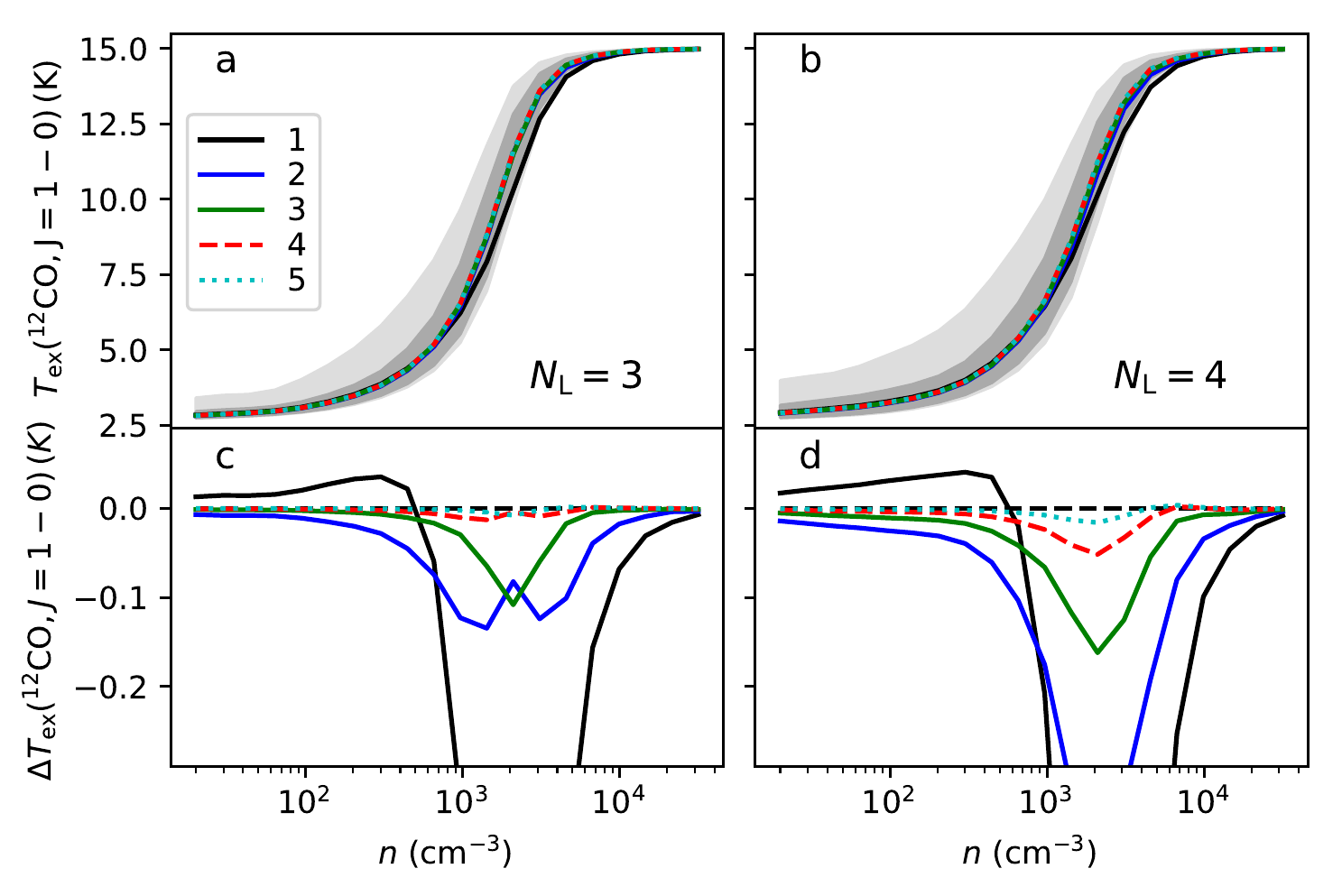}
\caption{
Average $^{12}$CO(1-0) excitation temperatures in different density bins in the
large-scale ISM simulation. The colours correspond to the situation after a given
number of iterations, as indicated in the legend. The left frames show results for
a model with a of 256$^3$ cell root grid and $N_{\rm L}=3$ and the right hand
frame for the same model with $N_{\rm L}=4$. For a given density, the grey bands
indicate the 1\%-99\% and 10\%-90\% intervals of the $T_{\rm ex}$ values.
The lower plots show the differences in the $T_{\rm ex}$ values between iterations
1-5 and the seventh iteration (an approximation of the fully converged solution).
The horizontal black dashed line corresponds to the $\Delta T_{\rm ex}=0$\,K level.
}
\label{fig:ROG_convergence}
\end{figure}

Figure~\ref{fig:ROG_convergence} shows the change of $^{12}$CO(1-0) excitation
temperature in different density bins and as a function of the number of
iterations. The calculations start with LTE conditions, with excitation
temperatures equal to the kinetic temperature, $T_{\rm ex}\equiv T_{\rm
kin}=15$\,K. The $T_{\rm ex}$ values are seen to converge in just a few
iterations. An accuracy of $\sim0.01$\,K is reached last at intermediate
densities, for $n({\rm H_2})$ a few times $10^3$\,cm$^{-3}$. This is because at
higher densities the transition is thermalised and the $T_{\rm ex}$ values
remain close to the original value. There relationship between volume density
and excitation temperature is not unique. For a given volume density, the
$T_{\rm ex}$ values can vary by up to a few degrees, depending on the local
environment, and show more scatter above than below the main trend.

We show in Fig.~\ref{fig:ROG_convergence} the results for two different model
discretisations with the number of octree levels $N_{\rm L}$=3 or 4.  The
calculations are seen to converge slightly slower in the $N_{\rm L}=4$ case. One
factor is the ALI acceleration, which is effective for optically thick cells.
When such a cell is split to sub-cells, some iterations are needed to find the
equilibrium state. Appendix~\ref{app:HCN} shows results from
additional convergence tests with the HCO+ and HCN molecules.

\begin{figure*}
\includegraphics[width=18cm]{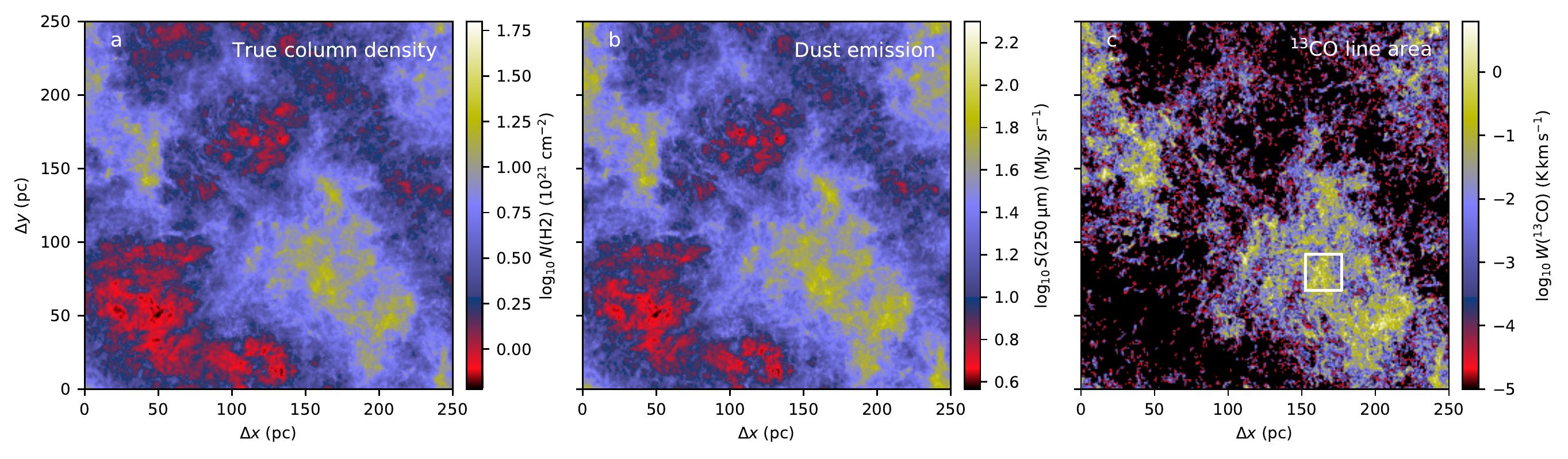}
\caption{
Comparison of dust and line emission for a large-scale ISM simulation. The frames
show the model column density (frame $a$), the computed 250\,$\mu$m surface
brightness of dust emission (frame $b$), and the$^{13}$CO(1-0) line area (frame
$c$). The white square in frame $c$ indicates an area selected for closer
inspection.
}
\label{fig:ROG_maps}
\end{figure*}

\subsubsection{Statistics of synthetic maps}  \label{sect:statistics}

We examine some basic statistics of the synthetic maps. This is done in part to
motivate the necessity for the time-consuming non-LTE calculations, instead of
using the more easily accessible quantities like the model column density or
the line emission calculated under the LTE assumption.

Figure~\ref{fig:ROG_maps} compares the computed line-area map $W(^{13}{\rm CO},
\/J=1-0)$ to the corresponding maps of the true column density and the surface
brightness of the 250\,$\mu$m dust emission. The dust emission is included as a
common tracer of ISM mass that also typically provides a higher dynamical
range. The details of the continuum calculations are given in
Appendix~\ref{app:continuum}. Figure~\ref{fig:ROG_maps} shows how the $^{13}$CO
emission is limited to the densest cloud regions, mainly because of the assumed
density dependence of the fractional abundances. The colour scale of the $W$
plot extends down to 10$^{-5}$\,K, although such low values are of course not
detectable in any practical line surveys. The noisy appearance of the line area
map is caused by abundance variations at small scales, although the variations
in the line excitation have a secondary but still important role. 

\begin{figure}
\includegraphics[width=8.8cm]{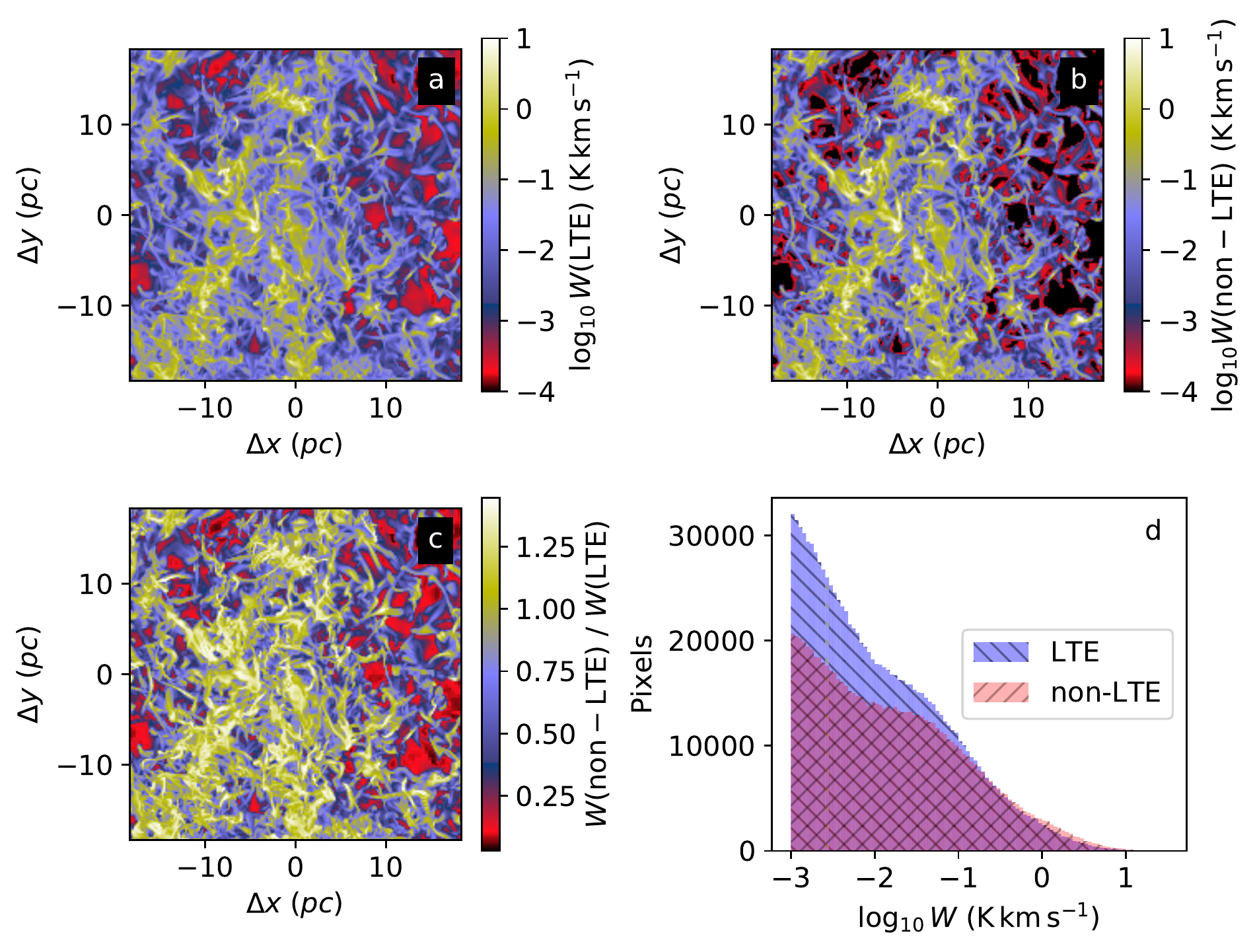}
\caption{
Comparison of LTE (frame $a$) and non-LTE (frame $b$) predictions of $W(^{13}{\rm
CO})$ for a region within the large-scale ISM simulation. The ratio of the LTE and
non-LTE maps is shown in frame $c$. The corresponding histograms of $W(^{13}{\rm
CO})$ values (above 10$^{-3}$\,K\,km\,s$^{-1}$) are shown in frame $d$. The
selected region is indicated in Fig.~\ref{fig:ROG_maps}$c$ with a white box.
}
\label{fig:W_13CO}
\end{figure}

The difference between the LTE and non-LTE emission is illustrated further in
Fig.~\ref{fig:W_13CO} for a selected small region. Size of the map pixels
corresponds to the smallest cell size of this $N_{\rm L}=4$ model, some
0.12\,pc, and is thus smaller than can be easily discerned in the plot. Both
calculations use the same model discretisation and $^{13}$CO abundances. In
addition to the lower average level of the non-LTE emission, the ratio of
non-LTE and LTE maps varies in a complex manner. In the histograms of
Fig.~\ref{fig:W_13CO}d, the difference increases towards lower column
densities. However, the non-LTE case has more pixels with very low emission and
its histogram peaks at densities far below the plotted range. In the more
practically observable range of $W(^{13}{\rm CO})\ga 10^{-2}$\,K, the
probability density distributions (PDFs) are more similar. It is also
noteworthy that the non-LTE predictions extend to higher intensities, in spite
of the average excitation temperature being much below the $T_{\rm kin}=15$\,K
value. This is caused by the higher optical depth of the $J=1-0$ transition,
when the higher excitation levels are less populated and the ratio of the $J=$0
and $J$=1 populations is higher.

\begin{figure}
\includegraphics[width=8.8cm]{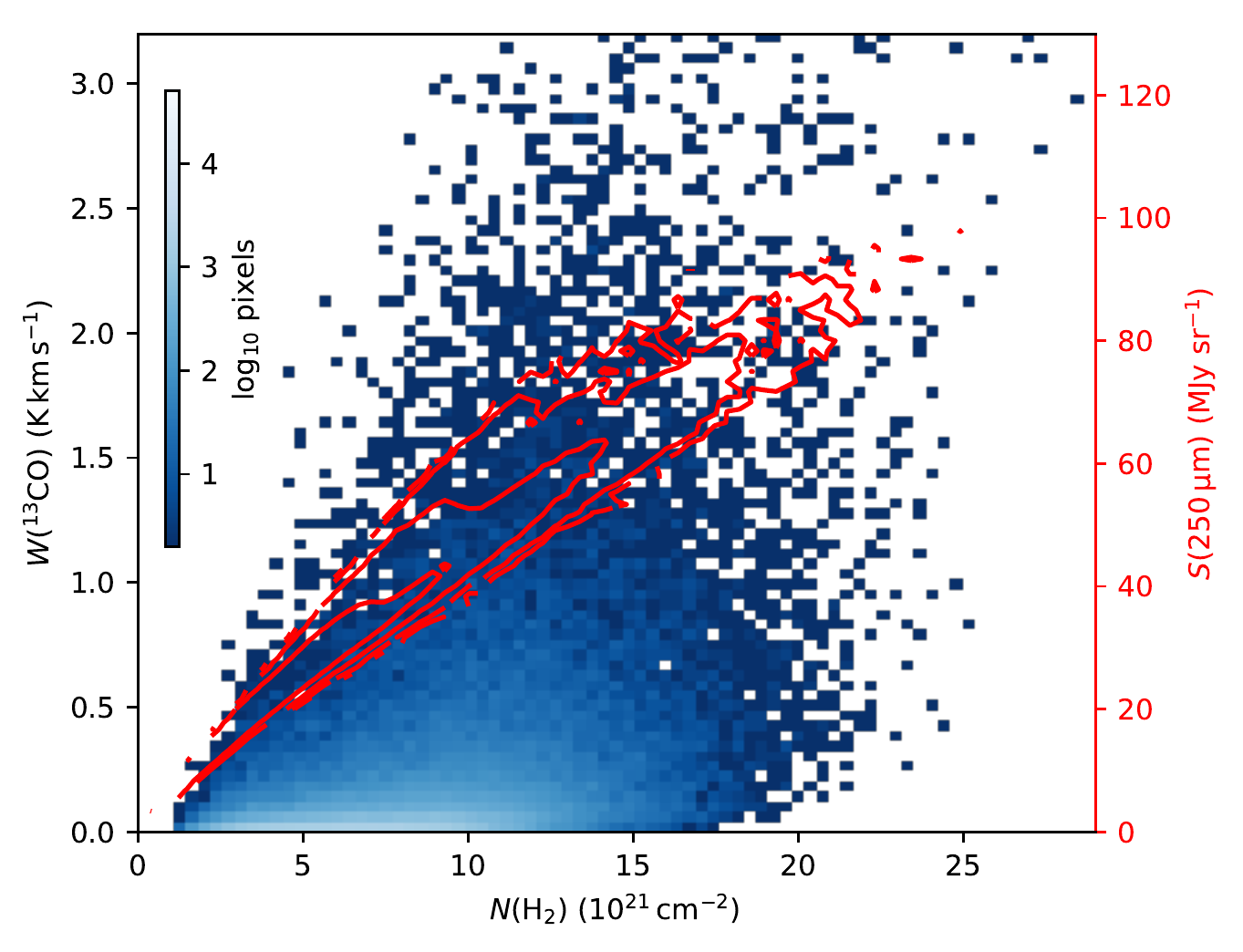}
\caption{
Dust and $^{13}$CO line emission as functions of column density of the large-scale
ISM simulation. The blue 2D histogram shows the distribution of the $^{13}$CO line
area vs. the true column density. The colour bar indicates the number of pixels in
the original line emission maps per 2D histogram bin. The corresponding
distribution of 250\,$\mu$m surface brightness vs. column density (right hand
axis) is shown with red contours. The contour levels correspond to
$\log_{10}n_{\rm pix}=0.5, 1, 2$, and 3, for the number image pixels $n_{\rm pix}$
per 2D histogram bin. The number of 2D bins is the same as for the 250\,$\mu$m
surface brightness histogram.
}
\label{fig:ROG_corr}
\end{figure}

Figure~\ref{fig:ROG_corr} shows the dust and line emission plotted against the
model column density. The dust emission shows good correlation with only some
saturation for the most optically thick LOS. In comparison, the scatter in the
line intensities is large. For a given column density, the line area can vary
by a significant factor, depending on the actual volume densities along the
LOS. The large variation also applies to the column density threshold above
which the line emission becomes detectable, $N({\rm H}_2)\sim (2-17) \times
10^{21}\,{\rm cm^{-2}}$. This scatter is again mainly due to the assumed
abundance variations, although the non-LTE excitation also plays an important
secondary role. Because of the ad-hoc nature of the assumed abundances,
Fig.~\ref{fig:ROG_corr} should be taken as a qualitative demonstration rather
than as a detailed prediction.

\begin{figure}
\includegraphics[width=8.8cm]{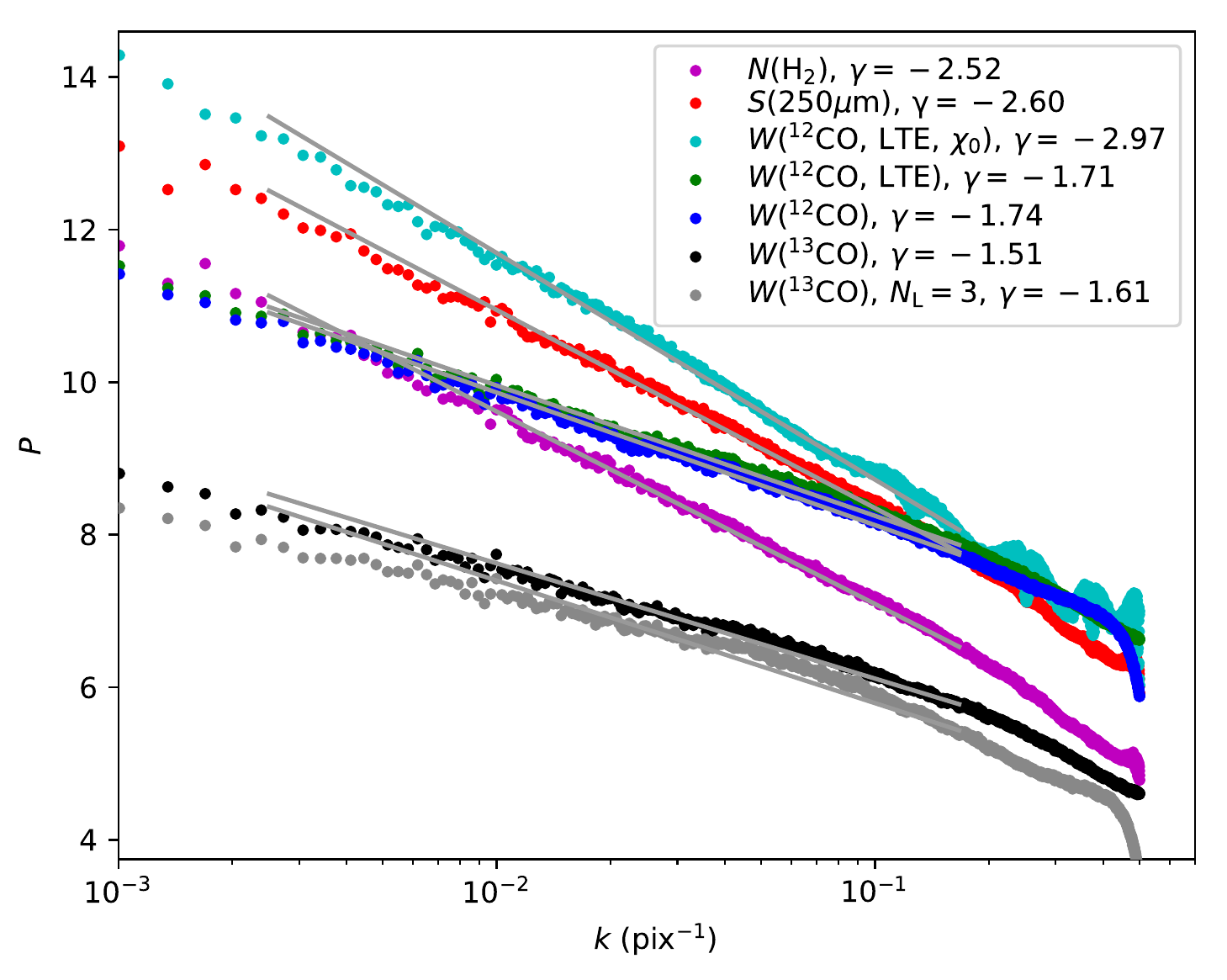}
\caption{
Power spectra from synthetic observations of the large-scale ISM simulation. The
power $P$ is calculated for the true column density, the 250\,$\mu$m dust surface
brightness, and for three versions of $^{12}$CO line-area maps. The $^{12}CO$
emission corresponds either to a constant abundance of $\chi_0=10^{-4}$ and LTE
conditions, or the Eq.~(\ref{eq:abundance}) abundances with LTE or non-LTE
calculations. The grey lines correspond to least-squares fits and extend over the
range of the fitted spatial frequencies.
}
\label{fig:PS}
\end{figure}

Figure~\ref{fig:PS} shows the power spectra calculated for different tracer maps.
These include line area maps for constant and density-dependent abundances
(Eq.~(\ref{eq:abundance})), both for LTE conditions and the full non-LTE
calculations with LOC. The power spectra were calculated with the TurbuStat
package \citep{Koch2019AJ}. The input maps have 2048$\times$2048 pixels and a
pixel size of 0.12\,pc, which corresponds to the smallest cell size of the models
(256$^3$ root grid and $N_{\rm L}=4$). The actual spatial resolution is lower over
most of the map area, but the discretisation is identical for all the compared
cases. The only exception is the $^{13}$CO map for the $N_{\rm L}=3$ model, which
is included as an example of a model with lower resolution. 

The dust surface brightness map and the constant-abundance LTE map of $^{12}$CO
line area have power spectra that are similar and even steeper than that of the
true column density. With the density-dependent abundances, the power is increased
at small scales, this resulting in much flatter power spectra. In the fits $P \sim
k^{\gamma}$ ($k$ being the spatial frequency), the slope of the power spectrum
rises above $\gamma=$-1.8. When the abundance distributions are identical, the
$^{12}$CO power spectra are slightly flatter for the non-LTE than for the LTE
case. The lower optical depth of the $^{13}$CO lines again increases the relative
power at the smallest scales, resulting in the largest $\gamma$ values. The
difference between the $N_{\rm L}$=3 and 4 discretisations has a smaller but still
a significant effect. The $N_{\rm }=3$ cell size corresponds to a spatial scale
outside the fitted range of $k$ values, but the reduction of peak intensities
(resulting from the averaging of cells in the $N_{\rm}=4\rightarrow3$ grid
transformation) is of course reflected to all values of $k$.

\section{Discussion and conclusions}  \label{sect:discussion}

The non-LTE, non-Monte-Carlo line transfer code LOC was shown to provide in the
selected 1D test problems (Sect.~\ref{sect:1d}) results comparable to those of
other RT programmes. The 3D version of LOC uses octree grids and is partly a
separate code base. In the tests of Sect.~\ref{sect:1D3D}, its results were
consistent with the 1D version. The 3D programme has also been compared with
the Monte Carlo code Cppsimu \citep{Juvela1997} (not discussed in this paper)
and those comparisons show a similar degree of consistency. The differences in
the results of different codes are also often connected with different
interpretations of the model setup rather than direct differences in the 
numerical methods themselves.

The parallelisation and the ability to run LOC on GPUs are important features.
After all, even on desktop systems, the number of parallel threads provided by
the hardware is approaching 10$^2$ on high end CPUs and 10$^4$ on high-end
GPUs. Therefore, the lack of parallelisation or poor scaling to higher core
counts would lead to unacceptable inefficiency, in terms of the run times and
the energy consumption. In the case of LOC, GPUs were found to provide a
typical speed-up of 2--10 over a multi-core CPU
(Appendix~\ref{app:performance}), although the numbers of course vary depending
on the actual model and the hardware used. The run times were found to be
directly proportional to the number of transitions and the number of cells. In
the case of very large models, the overhead from disc operations could decrease
the performance but that may be solved with more main memory. The run times
appeared to be rather insensitive to the number of velocity channels, but this
is at least partly due to the local emission/absorption profiles of the tested
models covering only a small fraction of the computed bandwidth. When the
number of channels is small, their impact on the run times is small because of
the relatively large effort spent on the basic ray tracing. When the number of
channels is larger, the run time should depend linearly on it. In details,
there may also be some more discrete behaviour because, on a GPU, a larger
number of channels (e.g. 32) are processed in parallel. The number of channels
should preferentially be a multiple of this local work group size (i.e. the
group of threads with synchronous execution).

The example of Sect.~\ref{sect:example} showed that relatively large models
with up to hundreds of millions of cells can be handled on modest desktop
computers and within a reasonable time. In Sect.~\ref{sect:example}, the
original model was reduced to 105 million cells and one RT iteration took less
than three hours. With more memory, larger modelling tasks could be tackled,
such as the full model that contained with 450 million cells (root grid 512$^3$
and seven octree levels, see Sect.~\ref{sect:example}). That would still be
within the reach of desktop computers and, with the most recent GPUs (compared
to the ones used in the tests), the run times would still be of similar order.
Conversely, RT calculations are needed also in the case of small 1D models.
When there are few observational constraints, simple 1D models are still a
common aid in the interpretation of observations. In fact, even much of the
theoretical work still relies on 1D calculations, e.g. in the study of
prestellar cores, where the spherical symmetry is a good approximation and the
emphasis is on chemical studies \citep{Vastel2018, Sipila2019}. Therefore, both
1D and 3D RT programs are still relevant. 

We showed in Sect.~\ref{sect:example} that differences in the synthetic
observations based on LTE and non-LTE calculations are clear even for the basic
statistics, such as the power spectra and the intensity PDFs. In the ISM
context, the non-LTE effects should be important for a wide range of numerical
and observational studies of clouds and cores
\citep{Walch2015,Padoan2016,Smith2020}, filaments and velocity-coherent
structures
\citep{Hacar2013,Hacar2018,Arzoumanian2013,Heigl2020,Chen2020,Clarke2020}, and
the large-scale velocity fields \citep{Burkhart2014,Hu2020}. These problems are
inherently 3D in nature, require the coverage of a large dynamical range, and
are natural applications of line RT modelling on hierarchical grids. At the
same time, RT is only one component of the problem and needs to be combined
with descriptions of the gas dynamics and chemistry. For example, the sample
calculations of Sect.~\ref{sect:example} made use of a simple density
dependence for the molecular abundances, but the real chemistry of the clouds
and especially the transition between the atomic and molecular phases is more
complex \citep{Glover_2010, Walch2015}. Tools like LOC are needed to quantify
the observable consequences of different model assumptions and to ultimately
test the models against real observations.

In the LOC results, the main errors are caused by the model discretisation
and the sampling of the radiation field with a finite set of rays. The latter
errors should decrease proportionally to $\sim N^{-1}$, where $N$ is the number
of rays. This is significantly faster than the $N^{-0.5}$ convergence of Monte
Carlo methods where, on the other hand, the errors are explicitly visible as
fluctuations between iterations. Some information of the sampling errors can be
extracted also in LOC calculations, for example by using different ray angles.
Changes in the spatial sampling are also possible but carry a large penalty
because the number of rays per root-grid surface element has to be either one
(the default value) or 4$^i$, with $i>1$. The noise in the level populations
translates to errors in the predicted line profiles. The line-of-sight
integration tends to smooth out the errors. Nevertheless, the relative errors of
the background-subtracted spectra (i.e. ON-OFF measurements) can again be
higher, if the lines are weak compared to the background continuum.

The computational performance of LOC is still far from the theoretical
peak performance, especially that of GPUs. The implementation of LOC as a
Python program with on-the-fly compiled kernels is well suited for further
development and testing of alternative schemes. When the device memory is
not an issue, an obviously faster alternative for the 3D computations would
be to process all transitions at the same time, instead of repeating the ray
tracing for each transition separately. Because the rate of convergence is
dependent on the local optical depths, some form of sub-iterations, where
level populations are updated more frequently in optically thick regions,
could result in significant reduction of the run times
\citep[cf.][]{Lunttila2012}. The number of energy levels that are actually
populated varies significantly from cell to cell. This was true for the
models discussed in Sect.~\ref{sect:example}, but would be much more evident
if the models included strong temperature variations, e.g. in the case of
embedded radiation sources. Further optimisations would thus be thus
possible by varying, cell by cell, the number of transitions for which RT
computations are actually performed and the number of energy levels for
which the level populations are stored.

LOC could also be extended with additional features. While it is possible to
have several collisional partners with spatially variable abundances in the 1D
models, the 3D version assumes their relative abundances to be constant (the
abundance of the studied molecule itself is always specified for each cell
separately). Continuum emission and absorption (e.g. from continuum RT
modelling) can not yet be considered in the 3D calculations. Similarly, while
LOC can model lines with hyperfine structure, the general (non-LTE) case of line
overlap is so far not included in the 3D version. Finally, if the spatial
discretisation is coarse, there might be some advantage from the use of
interpolated physical quantities. The use of velocity interpolation would
require sub-stepping, each small step with a different Doppler shift. Therefore,
further judicious refinement of the model grid might produce better results,
even considering the run times.

\bibliography{MJ_bib}

\begin{appendix}

\section{Computational performance of LOC}  \label{app:performance}

We examined the computational performance of LOC by comparing it to the Cppsimu
programme \citep{Juvela1997} and by measuring the run times as a function of
the model size. Cppsimu is a pure C-language implementation of Monte Carlo RT
but can be run with a regular, non-random set of rays. Thus, Cppsimu runs
without parallelisation provide a convenient point of reference, free from the
overheads that in the case of LOC result from the interpreted main programme
(in Python), the on-the-fly compilation of the kernels (a cost associated with
the use of OpenCL), the data transfers between the host and the device, and the
parallelisation.

\subsection{One-dimensional models}  \label{sect:time_1d}

When the Model 2b calculations of Sect.~\ref{sect:1D} were performed with the same
number of 75 iterations, using 256 rays and 64 velocity channels, the Cppsimu run
took on 3.1 seconds on a single CPU core, while LOC took 4.0 seconds on a CPU and
2.4 seconds on a GPU.\footnote{The one-dimensional models were run on a six-core
Intel i7-8700K CPU and NVidia GTX 1080 Ti GPU}. The sequential part of LOC before
the first call to the simulation kernel took only 0.2 seconds.

The LOC performance was modest because of the small amount of computations,
especially relative to the amount of data transferred between the host and the
device. When the number of rays is increased to 2048 and the number of velocity
channels to 256, the number of floating point operations should increase by a
factor of 64. For Cppsimu, linear scaling would predict a runtime of some 154
seconds while the actual run time was 354 seconds. This additional increase may be
attributed to the larger data volume (less efficient of use of memory caches) and
a lower average CPU clock frequency during the longer run. The LOC run on the CPU
took 183 seconds. This is half of the Cppsimu run time but not particularly good
given the availability of six CPU cores. On the other hand, the run time on the
GPU was 15.4 seconds, providing a speed-up by a factor of $\sim$23 over the
single-CPU-core Cppsimu run.

The results show that GPUs can provide significant advantages in RT calculations,
if the problem can make use of the available parallel processing elements. In the
above example, all 2048 rays could be processed in parallel because the GPU had an
even larger number of 2560 ``cores''. Although the 1D runs are never constrained
by the amount of available memory, LOC processes the transitions sequentially also
in the case of 1D models. By extending the parallelisation across both velocity
channels and transitions, the efficiency of the calculations for small models 
(smaller number of rays) could be improved.

\subsection{Three-dimensional Cartesian grids}  \label{sect:time_cart}

When the linear size of the root grid of the 3D models is at least some tens, the
parallelisation over simulated root-level rays should make use of most of the GPU
resources. The memory requirements are dictated by the large parameter arrays,
such as the density and velocity fields and the absorption counters, and less
memory is needed to support the concurrent processing of even thousands of rays.
When the number of threads is larger than the number of parallel processing
elements available in hardware, some of the idle time and overheads (e.g. due to
memory accesses) can also be amortised by fast context switching between the
threads. LOC uses different kernels for regular Cartesian grids and for octree
grids (discussed in the next section). On Cartesian grids, the ray-tracing
calculations have a small cost relative to actual the RT updates.

\begin{figure}
\includegraphics[width=8.4cm]{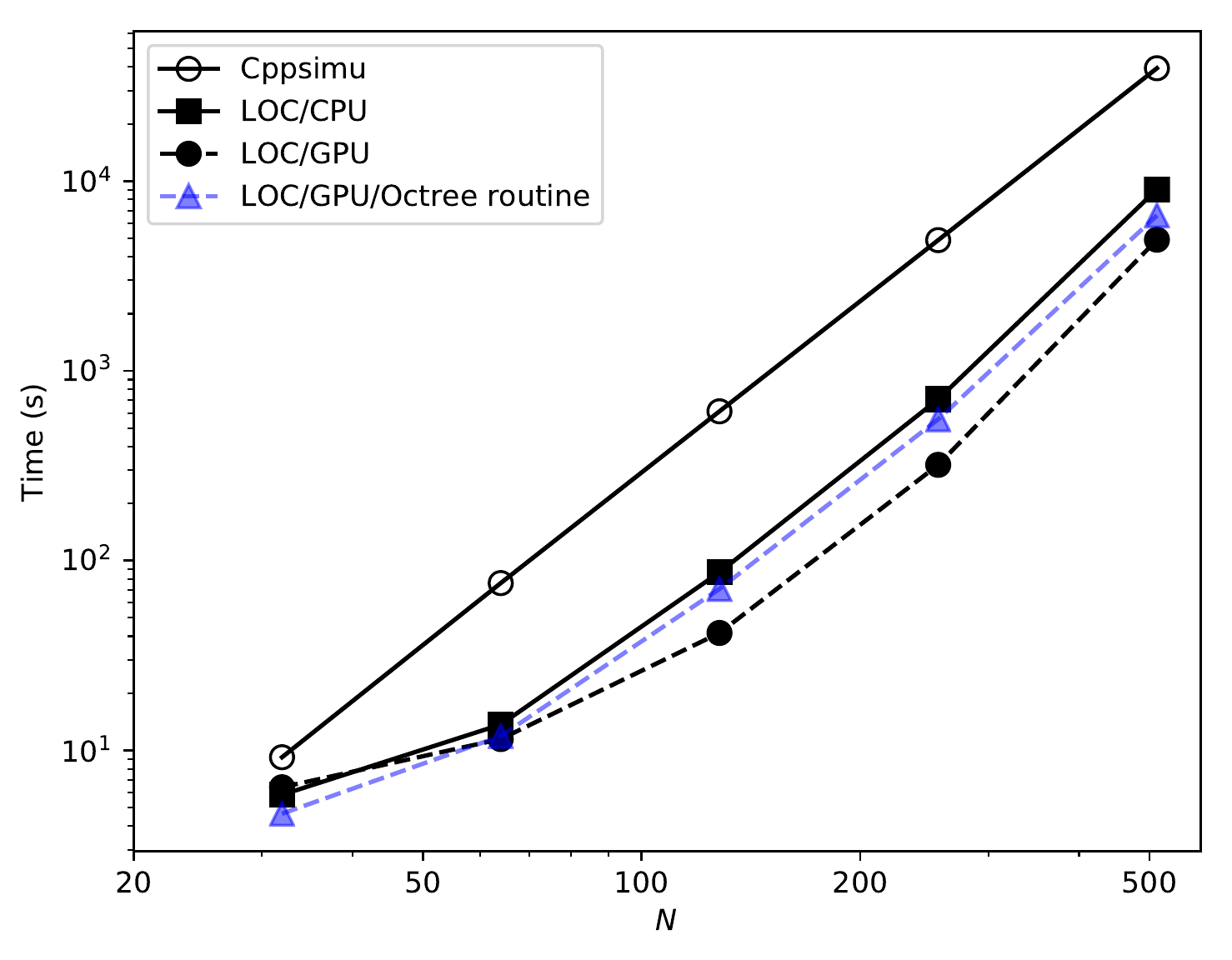}
\caption{
Comparison of Cppsimu (open circles) and LOC (filled black symbols) run times for
CO line calculations, using a 3D model discretised onto a regular Cartesian grid.
The x-axis gives the linear model size $N$, the models having $N^3$ cells, and the
y-axis the total run time, including the simulation of the radiation field and the
solving of the equilibrium equations. The figure also shows LOC run times when the
ray-tracing on the Cartesian grid is done with the more complex routines needed
for octree grids.
}
\label{fig:plot_time_cart}
\end{figure}

We compared the Cppsimu and LOC run times for a cloud model taken from
\citet{Padoan2016}. This is different from the snapshot discussed in
Sect.~\ref{sect:example} but also represents a (250\,pc)$^3$ volume of the ISM.
We performed RT calculations for the CO molecule, including the 10 lowest
energy levels and 512 velocity channels over a bandwidth of 50\,km\,s$^{-1}$
band. The model is discretised to $N^3$ cells with $N$ ranging from 32 to 512. 
Figure~\ref{fig:plot_time_cart} compares the run times for Cppsimu (CPU, single
core) and LOC on a CPU and a GPU\footnote{The CPU is a six-core Intel i7-8700K
processor and the GPU an external Radeon VII card, connected via a Thunderbolt
connection.}. The run times correspond to a single iteration that includes the
initial reading of the input files, the RT simulations, and the solving of the
equilibrium equations. Of these, the RT simulation is clearly the most time
consuming step. 

For $N=128-256$, LOC run on the six-core CPU provides a speed-up of $\sim$6.0
compared to the Cppsimu. This suggest good parallelisation for the six-core
CPU, although the number is of course affected by many implementation details
in the two programmes (and even the OpenCL version used). In fact, the speed-up
could have been expected to be smaller, because the average clock frequency is
lower during the multi-core LOC run (because of the thermal limits of the
processor and the laptop) and because Cppsimu simulates all transitions in one
go and thus does not repeat the ray tracing calculations for transition
separately. The speed-up provided by the GPU is $\sim$12.9 over the single-core
Cppsimu run and thus more than a factor of two over the LOC run on the six-core
CPU. The slightly smaller speed-up at $N=512^3$ may be affected by the
increased disc usage.

Both Cppsimu and LOC are optimised to limit the radiation field updates to the
velocity channels where the local absorption/emission line profile is
significantly above zero. In the turbulent ISM model, the local line profile is
typically much smaller than the full bandwidth of 50\,km\,s$^{-1}$. This means
that the geometrical ray-tracing has a relatively high cost relative to the
actual RT updates, in spite of the nominally large number of 512 velocity
channels. As was seen in the case of 1D models (Sect.~\ref{sect:time_1d}), the
relative performance of LOC improves if the number of floating point operations
is higher relative to the data volume, for example if the local
absorption/emission profiles covered a larger fraction of the bandwidth.

\subsection{Three-dimensional octree grids}  \label{sect:time_ot}

In the case of octree grids, the cost associated with the creation and storage
of rays and the more complex path calculations can be significant. In LOC, each
ray is processed by a single work group and the updates to the different
velocity channels are parallelised between the work items of that work group.
The creation and tracking of ray paths are delegated to a single work item,
which means that during those computations the other work items of the work
group remain idle. Therefore, the efficiency of parallelisation would again
increase if the number of velocity channels were larger (e.g. as in the case of
spectra with hyperfine structure).

Figure~\ref{fig:plot_time_cart} shows the run times also for the case where the
grid is Cartesian but we use these more complex ray-tracing routines needed for
octree grids. The overhead from the more complex path calculations is clear,
even when the grid itself does not yet have any refinement.

We examined further how the run times depend on the number of hierarchy levels and
the total number of cells in the octree grid.  We used a root grid of $64^3$ or
$128^3$ cells and 0-3 additional levels, up to the full spatial resolution of the
original 512$^3$ model cloud. The other run parameters are the same as in
Sect.~\ref{sect:time_cart}.

\begin{figure}
\includegraphics[width=8.8cm]{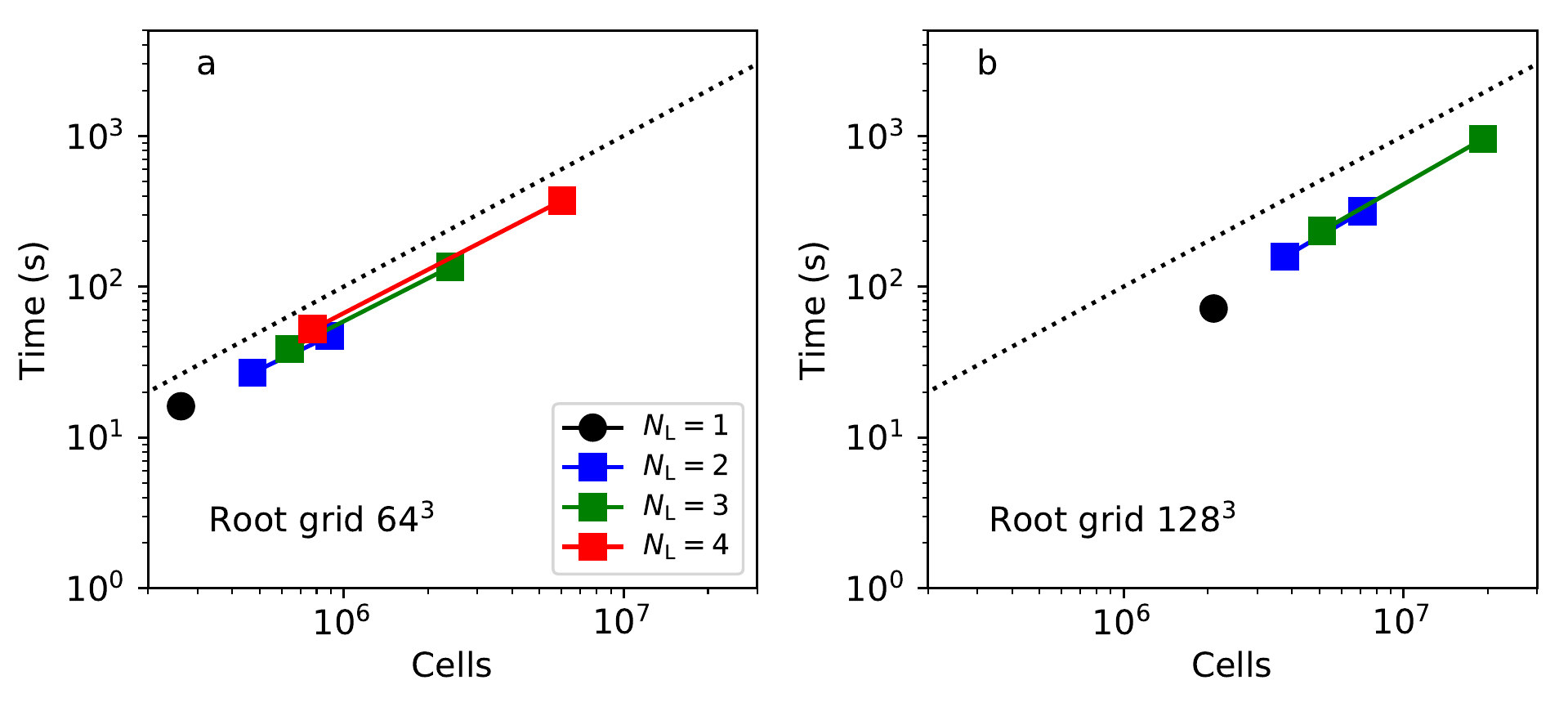}
\caption{
Comparison of LOC run times on a GPU for different octree grids. The octree
hierarchies have $N_{\rm L}=1-4$ levels and a root grid of $64^3$ (frame $a$)
or $128^3$ (frame $b$) cells. The solid lines connect two cases where either
10\% or 30\% of the cells of the previous level are refined. The dotted lines
indicate the slope of one-to-one scaling between the run time and the number
of cells.
}
\label{fig:plot_time_ot}
\end{figure}

Figure~\ref{fig:plot_time_ot} shows the run times as a function of the number
of cells in the model. Corresponding to each number of hierarchy levels $N_{\rm
L}$, there are two cases where either 10\% or 30\% of the cells of the previous
hierarchy level were refined, thus resulting in a different total number of
cells. When the percentage is 10\%, there is a nearly equal number of cells on
each refinement level (one tenth of the cells being split each to eight cells).
When the percentage is 30\%, most cells reside at higher hierarchy levels (i.e.
are smaller in physical size). The run times are seen to remain almost directly
proportional to the number of cells and there is little additional cost
associated with deeper hierarchies.

\section{Converge tests with HCO$^+$ and HCN} \label{app:HCN}

\begin{figure}
\includegraphics[width=8.8cm]{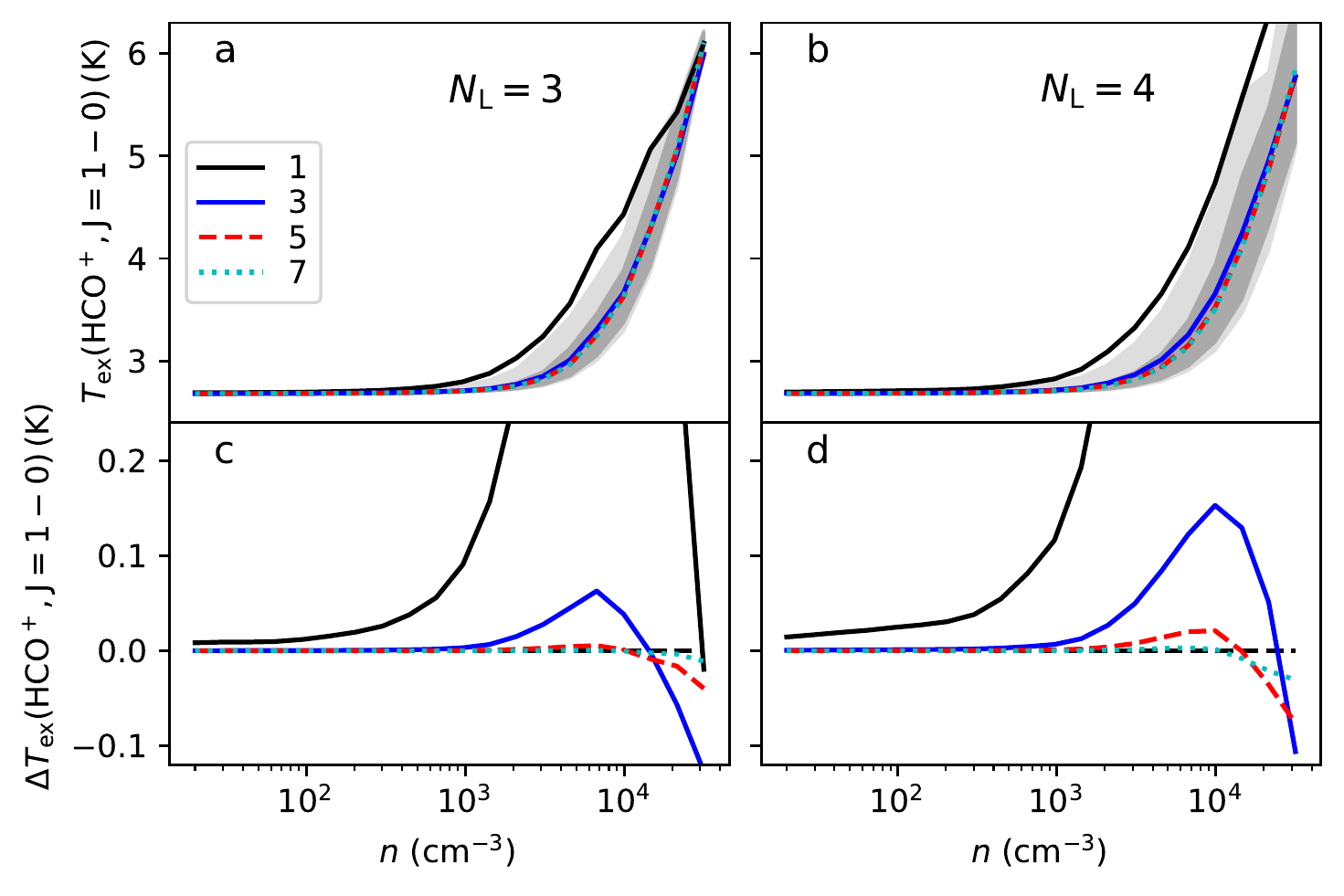}
\caption{
Average HCO$^+$(1-0) excitation temperature as a function of density in the
large-scale ISM simulation (cf. Fig.~\ref{fig:ROG_convergence}). The left and
right hand frames show the results for different spatial discretisations
($N_{\rm L}=$3, 4). The grey bands correspond to the 1\%-99\% and 10\%-90\%
$T_{\rm ex}$ intervals. The legend shows the colours that correspond to the
different number of iterations. 
The lower frames show the $T_{\rm ex}$ errors over iterations 1-7, estimated
relative to the final values after 12 iterations. The horizontal black dashed
line shows the $\Delta T_{\rm ex}=0$\,K level.
}
\label{fig:ROG_convergence_HCO+}
\end{figure}

In Sect.~\ref{sect:convergence}, the $^{12}$CO RT calculations indicated
that (when ALI is used) an increased spatial resolution may lead to some
decrease in the rate at which the level populations converge.
Figure~\ref{fig:ROG_convergence_HCO+} shows results for the same cloud model, in
the case of the HCO$^+$ molecule. The density dependence of the HCO$^+$
fractional abundances is similar to that used for $^{12}$CO but scaled to a
maximum value of $5\times 10^{-9}$. Similar abundances have been observed in
dense and translucent clouds, although the average value in the TMC 1 cloud is
slightly higher \citep{Fuente2019}.  However, in this test the main point is
that the optical depths are sufficiently large so that a number of iterations is
needed to reach the convergence. Figure~\ref{fig:ROG_convergence_HCO+} shows the
change of the average errors as a function of the number of iterations,
comparing the $N_{\rm L}=3$ and $N_{\rm L}=4$ discretisations. Although
HCO$^{+}$ remains sub-thermally populated, the largest $T_{\rm ex}$ residuals are
initially observed at intermediate densities. During later iterations (e.g.
iteration 7), the largest errors are at the largest densities. Similar
to the $^{12}$CO results (Fig.~\ref{fig:ROG_convergence}), the convergence is
again slower in the case of the finer spatial discretisation ($N_{\rm L}=4$).

\begin{figure}
\includegraphics[width=8.8cm]{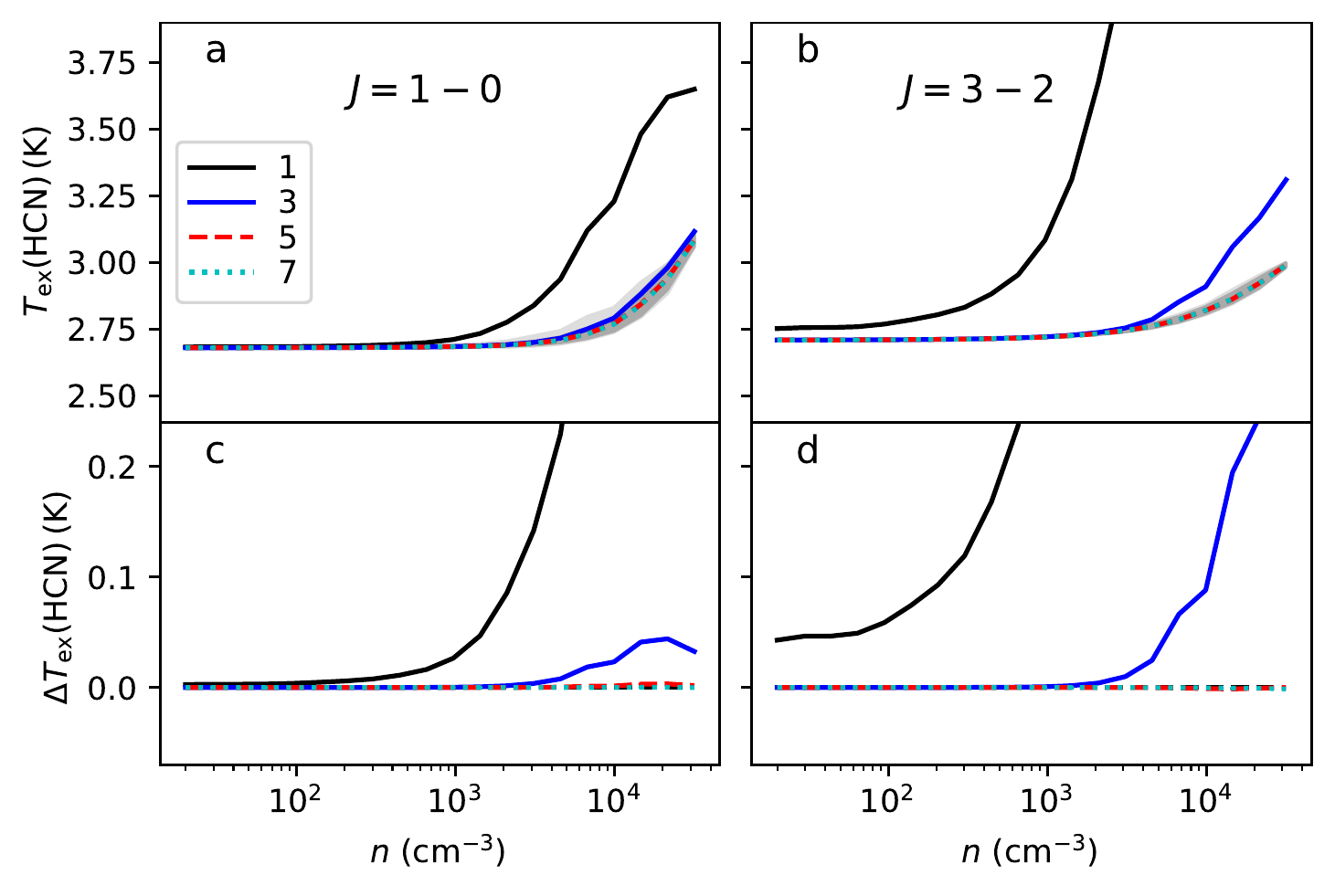}
\caption{
As Fig.~\ref{fig:ROG_convergence_HCO+} but showing convergence for the HCN
$J=1-0$ (left frames) and $J=3-2$ (right frames) transitions in the case of the
$N_{\rm L}=$3 model.
}
\label{fig:ROG_convergence_HCN}
\end{figure}

Figure~\ref{fig:ROG_convergence_HCN} shows a final test with the HCN
molecule. Here we use only the $N_{\rm L}$=3 model but plot the convergence for
both the $J=1-0$ and $J=3-2$ transitions. The hyperfine structure is taken into
account only for the $J=1-0$ transition, assuming LTE conditions between its
components. The maximum abundance of HCN was set to $5\times 10^{-9}$, which in
this case results in good convergence in less than ten iterations. The $T_{\rm
ex}$ residuals are initially larger for the higher transition, but the final
accuracy is similar for both transitions.

\section{Continuum RT calculations} \label{app:continuum}

In Sect.~\ref{sect:example}, the line calculations were compared with synthetic
dust emission maps. The continuum calculations use the same
octree-discretisation as the line calculations. The model is illuminated
externally by a radiation field that corresponds to conditions in the Solar
neighbourhood \citep{Mathis1983}. The dust properties are those of the $R_{\rm
V}=5.5$ model of \citet{Weingartner2001}, as implemented in the DustEM
package\footnote{https://www.ias.u-psud.fr/DUSTEM}.

The dust emission was calculated with the SOC programme
\citep{Juvela_2019_SOC}, assuming sub-millimetre emission from large grains at
an equilibrium temperature. Compared to line transfer, the continuum RT
calculations are faster, in spite of the use of Monte Carlo methods. By using
photon-splitting techniques, the noise of the dust temperature estimates
increases only linearly with the discretisation level, instead of the normal
factor-of-two increase per discretisation level. This is true as long as the
average optical depths are small and scatterings do not significantly reduce
the photon flux into the densest regions. This is true for the present models,
thus resulting in significant savings in the run times. The noise in the
computed dust temperatures was $\delta T_{\rm d}\sim 0.1$\,K at the highest
refinement level. The computed maps represent the surface brightness of dust
emission at the monochromatic wavelength of 250\,$\mu$m.

\end{appendix}

\end{document}